\documentclass[preprintnumbers, floatfix, preprintnumbers, letterpaper, twocolumn, superscriptaddress,nofootinbib]{revtex4-2}
\usepackage{graphicx}
\usepackage{microtype}
\usepackage{amsmath}
\usepackage{amssymb}
\usepackage{subfigure}
\usepackage{url}
\usepackage{hyperref}

\usepackage{xcolor}
\usepackage{color}
\usepackage{mathrsfs}
\usepackage{calrsfs}
\usepackage{amsfonts}
\usepackage{tabularx}
\usepackage{latexsym}
\usepackage{ragged2e}
\usepackage{epsfig}
\usepackage{textcomp}
\usepackage{float}
\newtheorem{theorem}{Theorem}
\newtheorem{corollary}{Corollary}[theorem]

\usepackage{caption}
\DeclareCaptionJustification{justified}{\leftskip=0pt \rightskip=0pt \parfillskip=0pt plus 1fil}
\definecolor{vividviolet}{rgb}{0.62, 0.0, 1.0}
\definecolor{amaranth}{rgb}{0.9, 0.17, 0.31}
\definecolor{palatinateblue}{rgb}{0.15, 0.23, 0.89}
\definecolor{brightpink}{rgb}{1.0, 0.0, 0.5}
\definecolor{cornflowerblue}{rgb}{0.39, 0.58, 0.93}
\definecolor{deepcarminepink}{rgb}{0.94, 0.19, 0.22}
\definecolor{radicalred}{rgb}{1.0, 0.21, 0.37}

\hypersetup{ linktoc=all,
	colorlinks, linkcolor={palatinateblue},
	citecolor={brightpink}, urlcolor={amaranth}
}

\graphicspath{{Images/}}

\def\sideremark#1{\ifvmode\leavevmode\fi\vadjust{\vbox to0pt{\vss
			\hbox to 0pt{\hskip\hsize\hskip1em
				\vbox{\hsize1.3cm\tiny\raggedright\pretolerance10000
					\noindent #1\hfill}\hss}\vbox to8pt{\vfil}\vss}}}%
\def\beq{\begin{equation}}
\def\eeq{\end{equation}}

\begin{document}
\title{Generalized Entropy Implies Varying-$G$:\\ Horizon Area Dependent Field Equations and Black Hole-Cosmology Coupling}

\author{Hengxin \surname{L\"u}}
\email{mx120220337@stu.yzu.edu.cn}
\affiliation{Center for Gravitation and Cosmology, College of Physical Science and Technology, Yangzhou University, \\180 Siwangting Road, Yangzhou City, Jiangsu Province  225002, China}

\author{Sofia Di Gennaro}
\email{sofia.digennarox@gmail.com}
	\affiliation{Center for Gravitation and Cosmology, College of Physical Science and Technology, Yangzhou University, \\180 Siwangting Road, Yangzhou City, Jiangsu Province  225002, China}

\author{Yen Chin \surname{Ong}}
\email{ycong@yzu.edu.cn}
\affiliation{Center for Gravitation and Cosmology, College of Physical Science and Technology, Yangzhou University, \\180 Siwangting Road, Yangzhou City, Jiangsu Province  225002, China}
\affiliation{Shanghai Frontier Science Center for Gravitational Wave Detection, School of Aeronautics and Astronautics, Shanghai Jiao Tong University, Shanghai 200240, China}

\begin{abstract}
When the Bekenstein-Hawking entropy is modified, ambiguity often arises concerning whether the Hawking temperature or the thermodynamic mass should be modified. The common practice, however, is to keep the black hole solution the same as that in general relativity. On the other hand, if Jacobson's method of deriving Einstein equations from thermodynamic is valid in the general settings, then given a generalized entropy one should first derive the corresponding modified gravity, and then look for the compatible black hole solution before investigating its thermodynamics. We comment on some properties and subtleties in this approach. In particular, we point out that generically generalized entropy would lead to a varying effective gravitational ``constant'' theory, in which $G_\text{eff}$ depends on the horizon area. We discuss in what ways such theories are discernible from general relativity despite its seemingly jarring differences, and how to make sense of area-dependent field equations. As a consequence we show that in the Jacobson's approach, the standard quantum gravitational logarithmic correction to Bekenstein-Hawking entropy is equivalent to a running gravitational ``constant''. A horizon area dependent $G_\text{eff}$ could also lead to a coupling between black hole masses and cosmological expansion, a scenario that has been studied recently in the literature, but so far lacks strong theoretical motivation. In the Tsallis case, we show that the thermodynamic mass for a Schwarzschild black hole is just a constant multiple of its ADM mass, which is considerably simpler than the approach not utilizing the Jacobson's method.
\end{abstract} 

\maketitle

\section{Introduction: Generalized Entropy and Modified Gravity}

In standard general relativity (GR), black holes have an entropy that is proportional to the horizon area. More precisely, in the units that\footnote{Here $k_B$ is the Boltzmann constant, $c$ is the speed of light in vacuum, and $\hbar$ is the reduced Planck constant.} $c=k_B=\hbar=1$, the Bekenstein-Hawking entropy is simply $S=A/4G$, the well-known ``area law''. 
What this entropy \emph{means}, however, remains unclear even after half a century. In the efforts to make sense of black hole entropy, 
various attempts have been made in the literature, which consider modifications to the Bekenstein-Hawking entropy, such as Barrow entropy \cite{2004.09444}, Tsallis(-Cirto) entropy \cite{Tsallis,Tsallis2}, Kaniadakis entropy \cite{0210467,0507311,2109.09181} and Sharma-Mittal entropy \cite{1802.07722}, just to name a few popular ones (entropy that generalizes these was constructed in \cite{2201.02424}). In this work, we are not necessarily advocating for the case of such generalizations, but instead we want to take this as an opportunity to study what may happen if we take these proposed modifications seriously. It is easy to say ``let's change the Bekenstein-Hawking entropy and see what happens to black hole physics'', but any modification will open up issues concerning consistency, and so it is important to examine generalized entropy more critically and sort out its various confusions in the literature.

One of the conceptual problems that arise from these generalized entropy is how to make sense of the corresponding first law of black hole thermodynamics\footnote{In cosmological applications, there are even more confusions \cite{2307.01768}.}. Let us consider only the uncharged and non-rotating case for simplicity. In GR the first law is straightforward\footnote{This does not mean that it is always simple. In asymptotically anti-de Sitter spacetimes, the concept of mass is notoriously tricky especially when the black hole is rotating \cite{2304.10290}.}: $dM=TdS$, where $M$ is the ADM mass of the black hole and $T=1/8\pi GM$ its Hawking temperature. However, once the entropy $S$ is modified to some new $\tilde{S}$, ambiguity follows. 
In the literature, one can typically find two approaches (see, e.g., the discussions in \cite{2106.00378}): 
\begin{itemize}
\item[(1)] Keep the Schwarzschild mass $M$ the same as thermodynamic mass. One should then define a temperature by $T_{\tilde{S}}:=\partial M/\partial \tilde{S}$.
\item[(2)] Keep the Hawking temperature, in which case the thermodynamic mass is changed into $E\neq M$, and the first law becomes $dE= T d\tilde{S}$.
\end{itemize}
In either case, one assumes that the black hole solution remains unchanged from that of GR\footnote{Another option is to demand that $M=E$ and $T$ remains the same as one would obtain from a \emph{modifed} solution, i.e., that theory is no longer GR, but not necessarily obtained via the Jacobson's method. See, for example, the discussions in \cite{2109.05315,2207.07905}.}. The first option appears to be somewhat problematic because at first glance there might be two temperatures -- one instinctual objection is that the standard Hawking temperature still exists since it is what one obtains from Wick rotating the metric in the usual manner and imposing regularity in the Euclidean period. It is not clear how to make sense of two temperatures; are they both physical? This criticism requires further examination, since the whole Wick rotation technique rests on the fact that in finite temperature quantum field theory, the Euclidean action $e^{-S}$ is formally identified with the Boltzmann factor $e^{-\beta H}$ in the partition function, which in turn implies that the standard Boltzmann-Gibbs distribution has been assumed. For example, if the Tsallis correction is introduced, we should consider the ``Tsallis factor'' \cite{0208205,0310413,Baranger} in place of the Boltzmann factor.
In other words, it is not at all obvious how the ``cyclic imaginary time = temperature'' identification, which is also tied to the regularization of the Euclidean manifold (to be free of conical singularity), is affected once a new statistical distribution is assumed. If everything is done consistently, perhaps there is only one well-defined temperature. This is an interesting question that nevertheless will not be pursued in this work, though it should be further investigated in view of interests in the field (see, e.g., \cite{2208.04473} for discussions concerning various ``temperatures''). However, it is worth mentioning that another common approach in the literature is to take the modified entropy, and compute the modified temperature via $dM=Td\tilde{S}$, and then find the metric that would give such an entropy upon Wick-rotation. In view of our discussion above, this may not be consistent because it seems to have implicitly assumed the usual Boltzmann-Gibbs entropy instead of a generalized one. In addition, the metric so obtained may not be unique, see \cite{2303.10719} for more discussions regarding similar approaches in the generalized uncertainty principle literature.

In the current manuscript,  the more crucial issue we are interested in is the following: applying generalized entropy directly to GR solutions may not be consistent, but can we do better?
We know from Jacobson's seminal work \cite{J} that one could start from thermodynamics considerations and obtain the Einstein field equations. In view of this, it can be argued that if the entropy is modified, we may no longer have the same black hole solution since the underlying theory is also modified. Indeed there are good motivations to study either modified gravity or generalized entropy. Yet, the relations between the two remain somewhat obscure and largely unexplored. There are some recent works in this direction, especially in the context of Barrow entropy \cite{2110.00059, 2207.09271, 2403.13687}. In \cite{2207.09271}, the case of Tsallis entropy was also discussed. However, there are still some subtleties that deserve further discussions and clarifications.

In the following, assuming that the parameter of the generalized entropy is constant, we will point out several interesting results:
\begin{itemize}
\item[(1)] If we use the Jacobson's approach, then the temperature is already chosen to be the standard Hawking one via the identification $T=\kappa/2\pi$, and so what should be changed is the thermodynamic energy. 
\item[(2)] The Jacobson's method generically yields field equations that depend on the area of the horizon. 
\item[(3)] If the modification of entropy is non-geometric in origin, i.e., the area of the horizon remains unmodified, then the Jacobson's method would give us a varying gravitational constant theory\footnote{The idea that the gravitational constant can vary has a long history, dating back at to Dirac's seminal paper in 1937 \cite{dirac}. See \cite{1009.5514} for a modern review.}. 
\item[(4)] Points (2) and (3) imply that the  varying gravitational constant is horizon area dependent.
\end{itemize}
We will then discuss whether such a theory is distinguishable from GR \emph{in practice}. (The answer is `yes', but not as straightforward as one might think as we need to be careful about the physical observables.) More importantly, we have to make sense of the consequences of having horizon area-dependent field equations, which renders the field equations quasi-local at best.

The structure of the manuscript is as follows.
In Sec.(\ref{sec2}) we start by reviewing the Jacobson's method for deriving the Einstein field equations and then apply it to generalized entropies, using Tsallis entropy as a concrete example. We comment on the subtleties in the Barrow case. We also state the general theorem that generalized entropy leads to area-dependence field equations.
Then in Sec.(\ref{3}) we derive the thermodynamic mass of Tsallis-corrected Schwarzschild black hole, noting the difference with the previous results in the literature. In Sec.(\ref{4}) we attempt to understand area dependent field equations. Then in Sec.(\ref{cosmo}) we briefly discuss some cosmological implications of such theories, including the possibility that black hole masses may be coupled with cosmological expansion. Finally, we conclude in Sec.(\ref{conclusion}) with more comments. We emphasize that we chose the Tsallis and Barrow form only for concreteness. At least for the context of this work we do not prioritize/advocate their validity more than any other generalized entropies. 

\section{Jacobson's Method Applied to Generalized Entropy}\label{sec2}
Let us briefly review Jacobson's method, which essentially makes use of the Clausius relation $\delta Q = T dS$.
First, we consider the heat flow across the horizon given by the matter flux
\begin{equation}
\delta Q =-\kappa \int_\mathcal{H} \lambda T_{ab} k^ak^b d\lambda dA,
\end{equation}
where $A$ is the area of the horizon $\mathcal{H} $ with $k^a$ denoting the tangent vector of the horizon generators, $\kappa$ the surface gravity, and $\lambda$ is an appropriate affine parameter. 
The Raychaudhuri equation gives the variation of the area as
\begin{equation}\delta A = \int_\mathcal{H}  \theta d\lambda d{A} =-\int_\mathcal{H}  \lambda R_{ab}k^ak^b d\lambda d{A},
\end{equation} 
where $\theta$ is the expansion of the horizon generators. 
Finally applying $\delta Q = T dS$, where $T=\kappa/2\pi$ is proportional to the surface gravity\footnote{In more general settings, the temperature is not necessary proportional to $\kappa$, see for example, the discussion in \cite{2307.16201}, in which the temperature is derived via the Euclidean grand canonical ensemble (Brown--York) procedure.}, we obtain
\begin{equation}\label{1}
-\kappa \int_\mathcal{H}  \lambda T_{ab} k^ak^b d\lambda d{A}= -\frac{\kappa}{2\pi} \int_\mathcal{H}  \lambda R_{ab}k^a k^b d\lambda d{S}.
\end{equation}
Then, with $S=A/4G$, this becomes
\begin{equation}
-\kappa \int_\mathcal{H}  \lambda T_{ab} k^ak^b d\lambda d{A}= -\frac{\kappa}{2\pi}\frac{1}{4G} \int_\mathcal{H}  \lambda R_{ab}k^a k^b d\lambda d{A}.
\end{equation}
This can only be valid if $T_{ab}k^ak^b=(1/8\pi G)R_{ab}k^ak^b$ for all null vector $k^a$, therefore we must have 
\begin{equation}\label{EFE}
(8\pi G)T_{ab}=R_{ab}+fg_{ab}
\end{equation}
for some function $f$. Finally, taking the covariant derivative and applying the contracted Bianchi identity 
\begin{equation}
\nabla^a R_{ab} = \frac{1}{2}\nabla_b R,
\end{equation}
fixes $f$ to be
\begin{equation}
f=\Lambda-\frac{R}{2},
\end{equation}
where $\Lambda$ is an integration constant that would serve as the cosmological constant once we substitute $f$ back into Eq.(\ref{EFE}) and obtain the Einstein field equations. 

Jacobson's method is rather general. For example, it is known that by including curvature correction (which amounts to non-equilibrium on the thermodynamic side) it is possible to derive $f(R)$ gravity \cite{0602001, 0909.4194}. In addition, by working with Wald entropy formula instead of the Bekenstein-Hawking entropy, one can obtain generic diffeomorphism-invariant theories of gravity \cite{0903.1254,0903.1176,0903.0823,1112.6215}. Having said that, one has to be careful of an implicit independent assumption in the Jacobson’s approach. Using the original Jacobson’s derivation as an example, GR was obtained from --- essentially --- the statement that $\delta E$ is directly proportional to $\delta A$. However, this is of course being computed in a certain frame. GR only follows if we impose that this calculation is true in \emph{any} frame. This was pointed out in \cite{1401.5262}. This is implicit since Jacobson's calculation assumed standard Riemannian/Lorentzian geometry (for example, utilizing the contracted Bianchi identity). This means that in principle one has to exercise caution on its validity when dealing with modified gravity theories; especially if diffeomorphism or Lorentz invariance is no longer valid.

Now let us move on to discuss generalized entropy.
To aid our discussions, let us look at a concrete example. (We remind the readers that the discussions below hold also for other generalized entropies; but it is easier to follow a concrete specific example rather than a general proof. We will state the general result later.)
Consider the black hole entropy under Tsallis entropy correction, which is motivated from non-extensive statistics:
\begin{equation}
S_T = \frac{A_0}{4G}\left(\frac{A}{A_0}\right)^\delta = \frac{A_0^{1-\delta}}{4G}A^\delta,
\end{equation} 
where $A_0$ is a constant and $\delta > 0$ is the Tsallis parameter (not to be confused with the variations in preceding equations).
Again, we emphasize that we are \emph{not} claming that black hole entropy is of the Tsallis form, but merely that this has been proposed in the literature, and we wish to clarify the implications for such a modification. Much of what follows would also hold for other generalized entropy; the Tsallis case is relatively easy to demonstrate.

Now, the LHS of Eq.(\ref{1}) remains \emph{unchanged} since Tsallis modification only changes the entropy expression; it does not modify the horizon area. The RHS, however, is modified. 
Assuming $\delta$ is constant, we end up with:
\begin{flalign}\label{2}
&-\kappa \int_\mathcal{H}  \lambda T_{ab} k^ak^b d\lambda d{A} = -\frac{\kappa}{2\pi} \frac{A_0^{1-\delta}}{4G}\int_\mathcal{H}  \lambda R_{ab}k^a k^b d\lambda d{A^\delta} \\ \notag
&=-\frac{\kappa}{2\pi} \frac{A_0^{1-\delta}}{4G}\int_\mathcal{H}  \lambda R_{ab}k^a k^b  \delta A^{\delta-1}  d\lambda dA.
\end{flalign}
That is,
\begin{equation}
-\kappa \int_\mathcal{H} \left[ T_{ab}  -   \frac{1}{8\pi G} {A_0^{1-\delta}}\delta A^{\delta-1}R_{ab} \right] \lambda   k^ak^b d\lambda d{A}= 0,
\end{equation}
which then implies 
\begin{equation}
8\pi G_\text{eff}T_{ab}=R_{ab}+fg_{ab},
\end{equation}
where the effective gravitational constant is 
\begin{equation}\label{tsallisG}
G_\text{eff}=\frac{G}{\delta}\left(\frac{A}{A_0}\right)^{1-\delta}.
\end{equation}
To get the modified version of the Einstein field equations, we have to take a covariant derivative and apply the contracted Bianchi identity. This results in
\begin{equation}
8\pi \nabla^a\left({G_\text{eff}}T_{ab}\right) = \frac{1}{2}\nabla_b R + \nabla_b f.
\end{equation}
At this stage, we have two choices: we can assume \cite{2012.05338} either that the conservation law is $\nabla^a\left({G_\text{eff}}T_{ab}\right)=0$, which is arguably more natural, or that it is still $\nabla^a\left({G}T_{ab}\right)=0$.
If we assume the former, then the resulting field equations read
\begin{equation}\label{GRe}
R_{ab}-\frac{1}{2}g_{ab}R + g_{ab} \Lambda = 8\pi G_\text{eff} T_{ab}.
\end{equation}
This is of course just GR but with $G$ replaced by $G_\text{eff}$. On the other hand, if we choose the latter, then we can obtain a considerably more complicated scenario 
\begin{equation}\label{odd}
8\pi G T_{ab} \nabla^a\left[{\frac{1}{\delta}\left(\frac{A}{A_0}\right)^{1-\delta}}\right] = \frac{1}{2}\nabla_b R + \nabla_b f.
\end{equation}
In the special case that the spacetime is stationary, this would still reduce to GR with $G_\text{eff}$. This is the case mentioned in \cite{2207.09271}. Note the subtle difference: if the conservation law is $\nabla^a\left({G_\text{eff}}T_{ab}\right)=0$, there is no need to impose stationary condition to obtain Eq.(\ref{GRe}).

Note that in the original work of Jacobson, he actually employed a \emph{local Rindler horizon} (only $dA$ appears in the derivation), instead of a black hole horizon. Of course the derivation also works with black hole horizon, but the crucial difference is that in the former, this holds \emph{locally} even if there is no global event horizon. Indeed, this is consistent with the field equation of general relativity being a set of PDE, which is of course, local. This is also consistent with previous findings that thermodynamic nature of the spacetime horizons is not restricted to the black holes, but applies also to the local causal horizons in the neighborhood of any point in the spacetime \cite{1304.2008}. (See, however, \cite{2207.04390}.)
In our case, once thermodynamics has been modified, however, since $G_\text{eff}$ depends on the \emph{area} of the horizon ($A$ appears in the result, not just $dA$), it is a quasi-local quantity. This seems radical and requires some serious discussions later.

At this point, we are ready to draw some conclusions. 
The most important observation is that the analysis above would follow through, \emph{mutatis mutandis}, as long as in Eq.(\ref{1}) the RHS is modified but the LHS is not, or more generally the LHS is modified in a different manner than the RHS. The only exception would be if both sides are modified in the same manner, as in the case of Barrow entropy analyzed in \cite{2207.09271}. In that case, the horizon becomes fractalized\footnote{Such a possibility was actually suggested earlier by Sorkin in \cite{9701056}, but he did not write down the form of the area law modification. Sorkin proposed that fractal structures due to horizon fluctuation may arise  on scales between the Planck scale $\ell_\text{P}$ and $(r_+\ell_\text{P}^2)^{1/3}$. One recognizes that the latter scale actually corresponds to the minimum error in the length measurement due to quantum gravity effect \cite{bf02717926,9406110,2405.16862}.} so its area is $A^{1+\Delta/2}$ for some $0\leqslant\Delta\leqslant 1$. Fundamentally the area law remains unchanged $S=\mathcal{A}/4G$, where now $\mathcal{A}=A^{1+\Delta/2}$. Thus one expects that Jacobson's method would simply give standard GR. However, in order to write down a Schwarzschild-like metric for which constant $r$ spacelike surfaces are still $S^2$, Ref.\cite{2207.09271} argues that we could view the area as unchanged, but that the entropy changes to Barrow entropy $S_B=A^{1+\Delta/2}/4G$. This would amount to an effective theory in which the fractal structure is not visible, but nevertheless it affects the thermodynamics. The LHS and RHS are modified in the same manner in Eq.(\ref{1}), but only \emph{re-interpreted} in the subsequent step. 
Namely, one starts with
\begin{equation}\label{barrow1}
\int_\mathcal{H} \lambda T_{ab} k^ak^b d\lambda dA^{1+\frac{\Delta}{2}} = \frac{\eta}{2\pi} \int_\mathcal{H} \lambda R_{ab}k^a k^b d\lambda dA^{1+\frac{\Delta}{2}},
\end{equation}
from which we get
\begin{equation}\label{f5}
\int_\mathcal{H} \lambda \left(T_{ab}-\frac{\eta}{2\pi}R_{ab}\right)\left(1+\frac{\Delta}{2}\right)A^\frac{\Delta}{2}k^ak^b d\lambda dA = 0,
\end{equation}
and consequently 
\begin{equation}\label{3}
\left(T_{ab}-\frac{1}{8\pi G}R_{ab}\right)\left(1+\frac{\Delta}{2}\right)A^{\frac{\Delta}{2}} = fg_{ab}.
\end{equation}
It is at this step that one re-interprets $A$ as the area of the black hole (instead of $A^{1+\frac{\Delta}{2}}$). 

The resulting field equations has no $G_\text{eff}$, but instead the cosmological constant is rescaled 
\begin{equation}
\Lambda \mapsto \frac{\Lambda}{\left(1+\frac{\Delta}{2}\right)A^{\frac{\Delta}{2}}}.
\end{equation}
Thus, despite the fact that both Tsallis and Barrow entropies have the form $A$ to some power, their modified gravity behaves quite differently. This is due to Barrow entropy arising from a geometric correction, whereas Tsallis entropy is the result of the modification of the underlying statistical mechanics distribution.

Even so, there are still some ambiguities as to \emph{when} we re-interpret the Barrow correction as merely an entropy correction that does not change the smooth horizon at the effective theory level. If we do so already at the beginning in Eq.(\ref{1}), 
then Eq.(\ref{barrow1}) would be changed to
\begin{equation}\label{barrow2}
\int_\mathcal{H} \lambda T_{ab} k^ak^b d\lambda dA = \frac{\eta}{2\pi} \int_\mathcal{H} \lambda R_{ab}k^a k^b d\lambda dA^{1+\frac{\Delta}{2}},
\end{equation}
in which the LHS is simply an integration with respect to $A$ instead of $A^{1+\frac{\Delta}{2}}$.
The result would then be similar to the Tsallis case, with $G_\text{eff}$ instead of a re-scaled $\Lambda$. Still, for both choices, in the vacuum case we will get just the Schwarzschild metric as in GR.

We see that in general, Jacobson's method leads to field equations that depend on the horizon area. 
In fact, the Bekenstein-Hawking entropy (up to a multiplicative constant) is unique in the sense that the field equations are area independent. To see this, we simply observe the following result:
\begin{theorem}\label{thm}
 If on the RHS of Eq.(\ref{1}) in the Jacobson's method we have an entropy correction to the horizon area law of the form $S=F(A)/4G$ for some differentiable function $F$, then $G_\text{eff}=G/F'(A)$. 
\end{theorem}
Here the prime denotes derivative with respect to the area. Then we immediately observe that:
\begin{corollary}{\textup{\textbf{(Uniqueness of Area Law)}}}
 $G_\text{eff}$ is independent of $A$ if and only if the area law holds, i.e., $S=\eta A$ for some constant $\eta$.
\end{corollary}
The proof is straightforward (follows the same calculation we did for the Tsallis example). We emphasize that this result is non-trivial, it is \emph{not} simply defining $G_\text{eff}$ by setting
$F(A)/4G=A/4G_\text{eff}$, in which case there would be no derivative of $F$.

In any case, this poses a serious question: does a generic gravity theory resulted from generalized entropy make sense only in spacetimes with a horizon? 
One way to avoid this situation is to allow the parameter in the generalized entropy to be a function. For example, in \cite{2403.13687}, when the Barrow parameter $\Delta$ is taken to be a function of $r$, the resulting modified field equations also have explicit area dependence, but with suitably chosen $\Delta$ the area dependence can be cancelled. This is similar\footnote{Though the argument for the running behavior therein is different: it is the requirement that for large black holes the leading term in the entropy, expanded as a series of $\Delta$, recovers the Bekenstein-Hawking form.} to the argument in \cite{2205.09311,2404.09278} in which Barrow entropy is assumed to run with energy scale and behaves at least as $\Delta \sim 1/A\ln A$.

\section{Example: Schwarzschild Black Hole with Tsallis Modification}\label{3}

Next, we note that in the derivation following Jacobson, the temperature is already chosen to be $T=\kappa/2\pi$, which is the standard Hawking expression, not the one defined by $\partial M/\partial S_T$, though the $\kappa$ here is the surface gravity $1/4G_\text{eff}M$ instead of $1/4GM$. This is because the $G$ in the Schwarzschild solution of GR should now be replaced by $G_\text{eff}$. The thermodynamic mass can be obtained following the method in \cite{2207.09271} or \cite{2106.00378}. Both of these, however, assume the standard Schwarzschild metric of GR\footnote{There is a typo in the coefficient of the thermodynamic mass of \cite{2106.00378} in the power of 2, upon fixing this it is the same as the result in \cite{2207.09271}.}, and so not compatible with the Jacobson's method. Unfortunately, despite already worked out $G_\text{eff}$, Ref.\cite{2207.09271} overlooked that the Schwarzschild solution should now be modified\footnote{One may think that since the Schwarzschild solution is a vacuum solution, it is not affected by $G \mapsto G_\text{eff}$ on the RHS of the field equations. This is unfortunately not true. In standard GR, $G$ in the Schwarzschild radius comes from considering the non-vacuum Newtonian limit (Poisson's equation), which would inherit the $G_\text{eff}$.}. 

To obtain the correct thermodynamic mass $E$, one begins with the first law
\begin{equation}
\int \frac{1}{T} dE = \int dS_T.
\end{equation}
The Schwarzschild equation now has horizon located at $r_h = 2G_\text{eff}M$. This leads to
\begin{equation}
dE= \frac{1}{32\pi G_\text{eff}^2 M}dA.
\end{equation}
We note that upon substituting $A=4\pi r_h^2$ into Eq.(\ref{tsallisG}),
\begin{equation}\label{G2}
G_\text{eff}= M^{\frac{2(\delta-1)}{1-2\delta}}f(\delta)^{\frac{1}{1-2\delta}},
\end{equation}
where we have defined
\begin{equation}
f(\delta):=\frac{(16\pi)^{\delta-1}A_0^{1-\delta}\delta}{G}.
\end{equation}
This function tends to $1/G$ as $\delta \to 1$ in the GR limit. (Though recall that we are assuming $\delta$ to be fixed, so $f$ is really a constant.)
With some simplifications we can obtain
\begin{equation}
dE=\frac{1}{G_\text{eff}}d(G_\text{eff}M),
\end{equation}
and finally upon integration the thermodynamic mass is given by a simple expression\footnote{This form of thermodynamic mass that only differs from the ADM mass by a constant (which in turn depends on the modification parameter) is previously known in the literature. For example, in bumblebee gravity the Schwarzschild-like solution \cite{1711.02273} has energy $E=\sqrt{1+l}M$, where $l$ is a parameter in the theory \cite{2401.15430}. Likewise, in Kalb-Ramond gravity, one finds \cite{2406.13461} $E=M/\sqrt{\ell+1}$ for some parameter $\ell$. Both $l$ and $\ell$ in these theories measure the degree of Lorentz violation.}
\begin{equation}\label{E}
E=\frac{1}{2\delta-1}M.
\end{equation}
The factor $f(\delta)$ does not appear in the final expression.
In the limit $\delta \to 1$, we recover $E=M$ for the standard GR result. This is a vast simplification compared to the results in \cite{2207.09271,2106.00378}, in which $E \propto M^{2\delta-1}$. We also note that the condition $\delta>1/2$ should be imposed to have a positive thermodynamic mass. Despite the differences in the approaches, this is the same bound obtained in \cite{2207.09271} from the consideration of Bekenstein bound. In the Tsallis-Jacobson case, the issue of Bekenstein bound violation \cite{2207.13652,2405.14799} never arises (as argued in \cite{2207.09271} we do not consider it as a violation if it is only off by a constant prefactor).

One of the advantages of the Jacobson's method is that we can check that with Eq.(\ref{G2}) for the expression of $G_\text{eff}$, the Tsallis entropy for fixed $M$ is monotonically decreasing towards 0 on the interval $\delta \in (1/2, \infty)$, although the expression $A^\delta$ might suggest that it diverges in the limit $\delta \to \infty$.
On the other hand, Tsallis entropy tends to a constant $A_0/4G$ as $\delta \to 0$, whereas $G_\text{eff}$ in Eq.(\ref{G2}) goes to zero. This is problematic since $G_\text{eff} \to 0$ should imply the lack of gravity, and so there is no black hole, yet the entropy is nonzero. Thankfully this unphysical situation is avoided due to the positive energy bound $\delta>1/2$.

Another interesting observation is as follows. In GR, the Smarr relation is $M=2TS$, whereas for a ``typical'' thermodynamic system one expects\footnote{Another way to obtain $E=TS$ is by introducing a pressure term that is associated with the ``thermodynamic force''  in GR  \cite{2309.04110}. In that case, one finds $E=M/2$.} $E=TS$ (a special case of the Euler equation). This can be obtained if we set $\delta=3/2$ in Eq.(\ref{E}). This would correspond to a ``volume law'' for the entropy (which is expected for ordinary systems like a box of gas) instead of an area law given by the Bekenstein-Hawking entropy, and thus serves as a consistency check of the result in Eq.(\ref{E}).

As we have repeatedly emphasized, the Tsallis case was used only as a concrete example. Thus,
interested readers may wish to follow this example to compute the thermodynamic energy for the Schwarzschild black hole in the other generalized entropies. In our opinion, a general consequence is that when dealing with thermodynamic calculations, e.g., phase transitions of a black hole spacetime with generalized entropy, it should be $E$ that is the relevant physical quantity, not $M$. We recommend that this be checked in the context of holography (gauge/gravity duality), to see if utilizing $E$ gives more sensible results on the field theory side than $M$.

\section{Distinguishability From General Relativity and Making Sense of Area Dependence}\label{4}

Imagine a planet Earth in an alternate universe governed by the field equations Eq.(\ref{GRe}), in which the area $A$ in $G_\text{eff}$ is the area of the cosmological apparent horizon. In such a world, local experiments would have always measured $G_\text{eff}$ instead of $G$. If $\delta$ is sufficiently close to 1, the effect of varying $G_\text{eff}$ would be difficult to detect, though its effect may show up in cosmological contexts (see Sec.(\ref{cosmo})), or astrophysical observations over a long time (for example, orbital changes in binary system whose masses are constant). Thus, physicists in such a world would eventually devise a theory of gravity identical to GR, and they would call $G_\text{eff}$ as simply some $\mathbb{G}$. The situation gets somewhat tricky when we consider black holes in such a universe. Since black holes have their own horizon, this means that the value of the $G_\text{eff}$ for black holes depends on their horizon area\footnote{The situation is actually even more complicated as one should somehow take into account the presence of multiple horizons. In GR we do not usually consider the expanding universe background when we discuss local black hole physics, so one might think that we can likewise ignore the cosmological horizon when we study black holes. However, this does not seem feasible because the cosmological horizon does imbue its $G_\text{eff}$ to all massive bodies in the universe. Therefore, the $G_\text{eff}$ for black holes must be a combination of the local and global values.}, and is distinct from the ``global'' cosmological $G_\text{eff}$. Nevertheless, their astronomers would not have realized this; to them, the effect of the different value of $G_\text{eff}$ is simply interpreted as the mass of the black hole. In other words, only $G_\text{eff}M$ is observable, which if they have never suspected that $\mathbb{G}$ can vary, then they would just interpret $G_\text{eff}M \equiv \mathbb{G}\mathcal{M}$ and deduce that the black hole has ADM mass $\mathcal{M}$, although it is really $M$. It would not make sense to fix $M$ to constrain $\delta$, as $M$ cannot be determined independently from astrophysical observations; it is only accessible from a God's eye view.  Note that the ``bare gravitational constant'' $G$ itself  in the defining expression of $G_\text{eff}$ is physically unobservable.

In the above scenario we have assumed that the cosmological horizon gives rise to $G_\text{eff}$ that governs gravity for all the objects inside the horizon\footnote{One may be tempted to think that horizon modifies $G$ into $G_\text{eff}$, and so ordinary objects without horizon would just feel $G$. However, this is not supported by the field equations. In the GR case, there is no $A$-dependence in the field equations so everything is governed by $G$. For the modified case, however, we cannot conclude that horizonless objects feel $G$.}. This seems to suggest a non-local effect at work, which some may deem to be not so desirable. Still, of course non-local extensions of GR is not rare either. In fact, since the size of the cosmological horizon is determined by the matter-energy content of the Universe, our claim here amounts to saying that the attraction of two particles depends on the total mass-energy content of the Universe, which is in the spirit of the ``Mach's principle'' (Mach tried to explain the inertial mass of any object as being induced by other masses in the Universe). One could also contemplate the possibility that away from the horizon, the value of $G_\text{eff}$ decreases as a function of proper distance. This does not change our above discussions by much, since the change in the value of $G_\text{eff}M$ of an astronomical body due to its local motion that changes its distance to the cosmological horizon would be imperceptible. Of course, careful observations may show some hints of a varying $G_\text{eff}$ as it values changes with cosmological expansion. There is a vast literature on constraining varying gravitational constants, see, e.g. \cite{2003.12832,2004.13976,2103.11157,2207.10674,2307.15382,2308.00233}.

The point is that \emph{we} could live in such a world already governed by Tsallis entropy without even realizing\footnote{While our approach is somewhat different, this discussion echoes the claim in \cite{2307.01768} that ``regardless of the type of entropy chosen on the cosmological horizon, when a thermodynamically consistent corresponding temperature is considered, all modified entropic force models are equivalent to and indistinguishable from the original entropic force models based on standard Bekenstein entropy and Hawking temperature ''. In our case the varying-$G$ effect may still be observable, but that alone cannot help us to decide whether we live in a modified entropy universe or other varying-$G$ universes.}. 
We would then treat $G_\text{eff}$ as a constant $\mathbb{G}$ and everything would just be GR. 
Note that the factor $1/\delta$ in $G_\text{eff}$ is quite subtle. We only obtained it by going through the Jacobson's derivation. If we had simply equate the actual entropy of the black hole in a Tsallis-Jacobson universe to the Bekenstein-Hawking entropy the naive inhabitants might prescribe to their black holes, i.e.,
\begin{equation}
\frac{A_0^{1-\delta}}{4G}A^\delta = \frac{A}{4\tilde{G}_\text{eff}},
\end{equation}
this would give instead,
\begin{equation}
\tilde{G}_\text{eff} = G\left(\frac{A}{A_0}\right)^{1-\delta},
\end{equation}
which is off by a $1/\delta$ factor compared to Eq.(\ref{tsallisG}). The reason was discussed right after Theorem \ref{thm}.

Similar conclusions can be reached for other non-geometric modifications to Bekenstein-Hawking entropy. Thus to summarize: if we take the influence of entropy on gravity seriously and accept that the Jacobson's method is universal, we are forced to accept that this generically leads to field equations that depend on the horizon area, and also generically the theory has an effective gravitational ``constant'' $G_\text{eff}=G_\text{eff}(A)$. Nevertheless inhabitants in such a universe likely would not realize this from local or even astrophysical measurements unless they are very careful (though there could be stronger hints coming from cosmology, see below). The area dependence in the field equations is somewhat bizarre, but can be dealt with in at least three manners depending on one's philosophy:
\begin{itemize}
 \item[(1)] If deemed undesirable, we could take this result as a \emph{reductio ad absurdum} and conclude that Jacobson's method -- for some reason -- only holds for entropy that satisfies the area law $S \propto A$. 
\item[(2)] If deemed undesirable, we can conclude that generalized entropy parameters cannot be constant, but rather must be chosen so as to cancel the area-dependence in the field equations.
\item[(3)] The derivation is sound and the result is what it is, \emph{fait accompli} -- just accept it. 
\end{itemize}

We prefer the last option, and would argue that it is actually not so strange. For example, following Theorem \ref{thm} in Sec.(II) that $S=F(A)/4 \Rightarrow G_\text{eff}=G/F'(A)$, the standard quantum gravity correction with a logarithmic term $S=A/4G + \text{const.} \ln(A)$ now implies that
\begin{equation}\label{glog}
G_\text{eff} = \frac{G}{1 + \frac{C_1}{A}},
\end{equation}
where $C_1$ is a constant. This is not too different from the running gravitational constant scenario one obtains from renormalization group argument in asymptotically safe gravity (ASG; see, for example, \cite{0610018,1807.10512,2212.09495,2302.04272}):
\begin{equation}
G(k)= \frac{G({k_0})}{ 1+ C_2k^2},
\end{equation}
where $k_0$ is a reference energy scale and $C_2$ another constant. 
We can assume $G(k_0)$ is the measured gravitational constant at our relatively low energy scale.
In the literature $k$ is often taken to be inversely proportional to a distance scale. In fact in \cite{2204.09892} (see also \cite{2308.16356}), the authors argued that $k$ is horizon area dependent and is explicitly given by $k = \text{const.}/\sqrt{A}$. This means $G(k)$ would recover the form in Eq.(\ref{glog}). Such a running behavior is more palatable since for large black holes, $G(k)$ essentially reduces to $G(k_0)$. The Jacobson's method therefore provides a link between the standard quantum gravitational correction of Bekenstein-Hawking entropy and the ASG approach.

We also remark that in the literature, \emph{distance}-dependent gravitational constant has been discussed \cite{Fujii,PhysRevD.9.850,PhysRevD.16.919,Hut}, e.g., a Yukawa potential type dependence in the context of the ``fifth force'':
\begin{equation}
G(r)=G_{\infty}\left(1+\alpha e^{-r/\lambda}\right),
\end{equation}
for some characteristic length scale $\lambda$. Here $\alpha$ is a constant or another function of $r$.
In comparison the area dependent version is arguably even less problematic (in terms of being well-defined), as the area of the event horizon of a black hole is coordinate independent. For the cosmological case, the apparent horizon is the most common choice when thermodynamic calculations are involved. This is because event horizons do not even exist in some Friedmann-Lemaître-Robertson-Walker (FLRW) spacetimes \cite{1106.4427}. The apparent horizon is observer dependent, but there is already a preferred frame in cosmology anyway. More specifically, the FLRW metric describes a homogeneous and isotropic universe only for observers like us who follow the ``Hubble flow''. The apparent horizon depends on the same FLRW cosmic time $t$. Note that there is no issue with ``general covariance''. For example, GR certainly satisfies general covariance as a theory, but one can construct \emph{useful} quantities that are coordinate-dependent such as light cones and apparent horizons. 
Furthermore, in $k=0$ FLRW geometry, the radius of the apparent horizon is just $R_H=1/H$, i.e. the Hubble radius, which is a physically relevant scale that can be computed from observation (of the Hubble constant). Some detailed clarifications can be found in \cite{1807.07587}, in which the name ``gravitational horizon'' is used to emphasize its physical nature (as opposed to a purely mathematical and arbitrary nature, as one may fear given the coordinate-dependent nature of the apparent horizon).

In addition, possible relation between mass and the cosmological horizon length scale has also been discussed in \cite{2307.06239}, in which the authors showed that different mass-horizon relation can recover the various generalized entropies, and in addition, that this allows the standard Hawking temperature to be used consistently via the generalized entropy. Possible connections between the effective gravitational constant and the curvature scale (Ricci scalar) has also been proposed in \cite{2312.02292}.

\section{Cosmological Implications}\label{cosmo}

We also note that, due to the reasons outlined in Footnote 11, the $G_\text{eff}$ for black holes should not simply be a function of its horizon area, but must also depend on the overall global, cosmological horizon area. But then, since $G_\text{eff}M$ is observed as $\mathbb{G}\mathcal{M}$, the overall effect is that the mass of black holes is coupled to cosmological expansion, a possibility recently investigated in the literature \cite{2212.06854,2302.07878,2306.08199,2307.10708,2307.02474,2312.12344,2405.12282}, essentially parametrized as:
\begin{equation}
\mathcal{M}(a)=\mathcal{M}\left(\frac{a}{a_i}\right)^k,
\end{equation}
where $a_i$ is the value of the scale factor when the black hole forms with initial mass $\mathcal{M}_i$, and the coupling parameter $k$ is bounded between the range $-3\leqslant k \leqslant 3$ from a causality constraint. The initial claim was that the best fit is $k\approx 3$ \cite{2212.06854,2302.07878}. However, recently Ref.\cite{2312.12344} argued that current estimates of the supermassive black hole mass density based on the black hole mass -- bulge mass relation probably exclude $k=3$, but left the possibility of $k>2$ opens. The problem, however, is that in our case, we do not actually know how the $G_\text{eff}$ due to black hole horizon combines with the $G_\text{eff}=\mathbb{G}$ of the cosmological horizon, and thus we do not know how $\mathcal{M}$ looks like as a function of $a$. Nevertheless, perhaps with some assumptions, reasonable models can be constructed for future study. The point is that, Jacobson's method applied to Tsallis entropy (and other generalized entropy) provides a \emph{theoretical basis} for the possibility of the coupling between black hole masses and cosmological expansion, which was hitherto not available (the possible coupling was proposed phenomenologically without a theoretical motivation). In our picture, this coupling
is simply due to the different $G_\text{eff}$'s, it does not necessary imply that black holes are a ``source'' for dark energy as originally claimed \cite{2302.07878}.

Although a full analysis of cosmology in the Tsallis-Jacobson approach is beyond the scope of this work. A few interesting observations are worth a quick mention. Let us consider a Friedmann–Lemaître–Robertson–Walker universe with flat spatial section. Following \cite{0807.1232}, we apply the Tsallis correction to the apparent horizon\footnote{As shown in \cite{0704.0793}, the applicability of the first law of thermodynamics to the apparent horizon has deep physical and geometric roots, which essentially rely on a linear mass-to-horizon relation, which remains valid in our case.  In general however, one should be very careful to guarantee consistency if other forms of entropy and/or temperature are used. See \cite{2307.06239} for more discussions.} at $R=1/H$, where $H=\dot{a}/a$ is the Hubble parameter. Applying the conservation law $\nabla^a\left({G_\text{eff}}T_{ab}\right)=0$ yields a modified fluid equation\footnote{If so desired, $\ddot{a}/a$ can be substituted with the acceleration equation, so that the fluid equation is expressed entirely in terms of the Hubble parameter $H$.}
\begin{equation}
\dot{\rho} + 3H(\rho+p) + 2(1-\delta)\rho H\left(1-\frac{1}{H^2}\frac{\ddot{a}}{a}\right)=0.
\end{equation}
This gives rise to certain complications. For example, we can no longer work with $\rho(t) \propto 1/a(t)^3$ for pressureless matter as per usual. Unfortunately, the acceleration equation is the same as in GR, but with $G$ replaced by $G_\text{eff}$, so there is no chance for accelerated expansion with just normal matter. The Tsallis-Jacobson cosmology thus differs from the earlier Tsallis cosmology approach in \cite{1806.03996}. What is potentially interesting is that 
\begin{equation}\label{varyG}
\frac{\dot{G}_\text{eff}}{G_\text{eff}}=(1-\delta)2H\left(1-\frac{1}{H^2}\frac{\ddot{a}}{a}\right).
\end{equation}
In terms of the modified fluid equation, we can write
\begin{equation}
\dot{\rho} + 3H(\rho+p)=-\rho\frac{\dot{G}_\text{eff}}{G_\text{eff}},
\end{equation}
which is indeed the known general behavior for varying Newton's constant cosmology, see, e.g., \cite{1910.08325}. Eq.(\ref{varyG}) is interesting as it vanishes for a pure de Sitter spacetime with scale factor $a(t)=a_0 e^{Ht}$, where $H$ is a constant. As long as the expansion rate is not faster than the pure de Sitter case, and if $\delta=1-\varepsilon < 1$, then $\dot{G}_\text{eff}>0$. We can therefore expect early Universe to have a smaller value of effective Newton's constant, which could help to ameliorate the arrow of time problem\footnote{This also works for the $G_\text{eff}$ in Eq.(\ref{glog}), namely we observe that $G_\text{eff} \to 0$ when $A \to 0$ (or $k\to \infty$ in the ASG approach; i.e. gravity weakens at high energy scale).} \cite{0911.0693,1810.06522}, and may also help to relax the Hubble tension and $S_8$ tension \cite{2102.06012,2201.11623} (see, however, \cite{2112.14173}). Then, as the Universe becomes asymptotically de-Sitter (as indicated by observations), $\dot{G}_\text{eff}\sim 0$ and we have an almost constant ${G}_\text{eff}$. In other words, curiously, the growth of $G_\text{eff}$ is kept in check by the late time quasi-de Sitter geometry. This connection between ${G}_\text{eff}$ and de Sitter geometry may tie in with the proposal in \cite{1910.08325} that varying-$G$ may provide an effective ``cosmological constant''.
Despite all these prospects, we do expect that observational constraints would likely tightly constrain $\delta$ to be very close to 1 as the correction term cannot be too large (especially during the early epochs when the universe is not close to de-Sitter-like). But again, in general, one must be careful about what is exactly observable \cite{1004.2066}.

Another prospect for a horizon-dependent $G_\text{eff}$, as opposed to other varying-$G$ scenarios, is that we should expect it to depend on all the horizons in the spacetime in a complicated manner. This is both an advantage and at the same time, a challenge. It is advantageous in the sense that this seems to suggest that the value of $G_\text{eff}$ can change abruptly if the number of horizons suddenly increases, say, during an era in which black holes are copiously produced. A ``rapid transition'' of the gravitational constant has indeed been considered in cosmology \cite{2102.06012,2202.09356,2306.05450}. While it appears that \emph{when} such a transition may have happened is still debatable, it would be interesting to investigate if such a transition may coincide with black hole formation rate that is higher than usual, or changes in the cosmological apparent horizon. In particular, in \cite{2202.09356} the authors argued that there is a shift in the effective Planck mass around the recombination era. This is interesting since the increase in matter density during recombination could reduce the Hubble constant \cite{2105.03704}. If true, this is equivalent to increasing the size of the cosmological horizon (whose radius is the inverse of the Hubble constant), and hence $G_\text{eff}$. Of course, this qualitative conjecture would require a more careful study. The main challenge lies in the multi-horizon case, as we do not know how the global $G_\text{eff}$ is determined from the ``interactions'' between the horizons. However, if black hole-cosmology coupling is real, we could use observational data as a guide to further understand the possible form of the interactions.

\section{Conclusion: Entropy Fixes the Gravity Theory}\label{conclusion}
The link between gravity and entropy has been widely discussed, but many facets remain unclear and mysterious. 
In this work, we have taken a closer look at the influences of generalized entropy on black holes and cosmology.
We argued that if we first consider the effect of generalized entropy on the gravitational theory itself via the Jacobson's method, instead of taking GR solution and impose the entropy correction \emph{a posteriori}, the results are quite different. The former approach is arguably more consistent, but is not without problems. Namely, the field equations are now at best quasi-local, being dependent on the area of the horizon. How does one make sense of spacetimes with no horizon in such a theory? If one naively takes absence of horizon as the same as $A \to 0$, this would seem to imply that gravity requires horizon -- without horizon, $G_\text{eff}$ is zero. Of course, area dependence can also be interpreted as a mass-energy dependence (for a Schwarzschild black hole, the relation between area and mass is straightforward; for apparent horizon in FLRW cosmology, it is implicit via the Friedmann equation). In this sense, the situation is similar to the Brans-Dicke theory in which the effective gravitational constant $G$ is replaced by the inverse of a scalar field: $G_\text{eff}=\phi^{-1}$, wherein the scalar field itself is related to the matter field by $\Box \phi = 8\pi/(3+2\omega)T$, where $\omega$ is the Brans-Dicke coupling constant and $T=T^a_{~a}$ is the trace of the energy-momentum tensor. Our situation is still somewhat different, however, as vacuum solutions can affect $A$, and non-vacuum solutions with no horizons \emph{do not}. The roles of an effective gravitational constant have been explored in the literature (see, e.g., \cite{0712.3206,1005.2327}), but in view of our findings, there are clearly still rooms for future investigations.

The Jacobson's approach would also be more restrictive -- once a generalized entropy is given, there is a modified gravity that corresponds to it. This means we can no longer consider, for example, Barrow or Tsallis correction on top of another modified gravity theory. 
Such a restriction is good as it avoids proliferation of generalizations in the absence of guiding principles.
Note that we are not claiming that there is a one-to-one map between generalized entropy and modified gravity; as not all modified gravity theories can be derived via the Jacobson's method. 

There are, in any case, interesting physics that follow from the Jacobson's approach, as we have seen from our example (Tsallis-Jacobson gravity). Therefore, we would suggest that further research should be carried out to examine if other generalized entropy can lead to new features once their corresponding modified gravity is derived, especially in cosmological contexts. For example,
we expect that ``Kaniadakis-Jacobson cosmology'' would be quite different from Kaniadakis cosmology \cite{2406.11373}.
Finally we should emphasize that in this work, we have only considered the simplest scenario in which the generalized entropy parameter is assumed to be constant. However, there are good reasons to suspect that it would run with energy scale, just like in the case of Barrow entropy \cite{2205.09311,2404.09278,2403.13687}. What happens then to the corresponding modified gravity theory? Clearly more research is needed.

To conclude, the relationship between generalized entropy and modified gravity calls for more studies, as there is a potential that this might lead to interesting theories or models that can tie in with other proposals in the literature. At the very least, it forces us to think deeper about some very important questions: how to consistently modify gravity and thermodynamics? Is GR unique in some thermodynamical sense?

\end{document}